\newcommand{\diracslash}[1]{#1\llap{/\kern2pt}}

\newcommand{\be}{\begin{equation}}
	\newcommand{\ee}{\end{equation}}
\newcommand{\bea}{\begin{eqnarray}}
	\newcommand{\eea}{\end{eqnarray}}
\newcommand{\ba}[1]{\begin{array}{#1}}
	\newcommand{\ea}{\end{array}}

\newcommand{\bt}{\begin{tabular}}
	\newcommand{\et}{\end{tabular}}
\newcommand{\Tr}{{\rm Tr}}

\newcommand{\ovl}{\overline}

\newcommand{\beas}{\begin{eqnarray*}}
	\newcommand{\eeas}{\end{eqnarray*}}

\documentclass[preprint,prd,aps,amssymb,amsmath
,floats,nofootinbib,floatfix]{revtex4}

\DeclareSymbolFont{rsfs}{U}{rsfs}{m}{n}
\DeclareSymbolFontAlphabet{\mathrsfs}{rsfs}

\usepackage{graphicx}
\usepackage{multirow}
\usepackage{graphicx,epstopdf}
\usepackage[autostyle]{csquotes}

\usepackage{graphicx}
\usepackage{float}

\usepackage[utf8]{inputenc}
\usepackage[english]{babel}
\usepackage{hyperref}
\hypersetup{
	colorlinks=true,
	linkcolor=blue,
	filecolor=magenta,     
	citecolor=blue, 
	urlcolor=blue,
}

\usepackage{cleveref}

\begin{document}

	\title{$\eta$ mesons in hot and dense asymmetric nuclear matter} 
	
	\author{Rajesh Kumar}
	\email{rajesh.sism@gmail.com}

	\author{Arvind Kumar}
	\email{kumara@nitj.ac.in}
	\affiliation{Department of Physics, Dr. B R Ambedkar National Institute of Technology Jalandhar, 
		Jalandhar -- 144011,Punjab, India}
	%

	\def\be{\begin{equation}}
		\def\ee{\end{equation}}
	\def\bearr{\begin{eqnarray}}
		\def\eearr{\end{eqnarray}}
	\def\zbf#1{{\bf {#1}}}
	\def\bfm#1{\mbox{\boldmath $#1$}}
	\def\hf{\frac{1}{2}}
	\def\kp{\zbf k+\frac{\zbf q}{2}}
	\def\km{-\zbf k+\frac{\zbf q}{2}}
	\def\hwo{\hat\omega_1}
	\def\hwt{\hat\omega_2}

	\begin{abstract}
		We study the $\eta N$ interactions in the hot and dense isospin asymmetric nuclear matter using two different approaches. In the first approach, the in-medium mass and optical potential of $\eta$-meson have been calculated in the chiral SU(3) model,  considering the effect of explicit symmetry breaking term and range terms in the $\eta N$ interaction Lagrangian density. In the second scenario,  the conjunction of chiral perturbation theory and chiral SU(3) model is employed. In this case, the next-to-leading order $\eta N$ interactions  are evaluated from the chiral perturbation theory (ChPT), and the in-medium contribution of scalar densities are taken as input from chiral SU(3) model.  We observe a larger negative mass-shift in the ChPT+chiral model approach compared to the chiral SU(3) model alone as a function of nuclear density. Moreover, the increase in the asymmetry and temperature cause a decrease in the magnitude of mass-shift. We have also studied the impact of $\eta N$ scattering length $a^{\eta N}$ on the $\eta$ meson mass $m^*_\eta$ and observed that the  $m^*_\eta$ decrease more for increasing the value of scattering length.  

	\end{abstract}

	\maketitle

	\maketitle
	
	\section{Introduction}
	\label{intro}
	
	The meson-baryons interactions are very imperative topic of research to study the physics of non-perturbative QCD regime \cite{Tolos2020,Papazoglou1999,Hayashigaki2000,Jenkins1991,Kaplan1986,Tolos2004,Tolos2006,Tolos2008,Cieply2014,Zhong2006,Waas1997,Vogt2007}. The heavy ion-collisions (HICs) are used to study the strong-interaction physics by colliding high energy particles. As a byproduct of the collision, the  Quark Gluon Plasma (QGP) appears under the utmost conditions of density and temperature \cite{Vogt2007}. Afterward with the expansion of fireball the QGP cools down and changes its phase to the hadronic matter through hadronization process \cite{Vogt2007}. These two regimes $i.e.$ QGP phase and hadronic phase have  different characterization of the respective  medium. For example, in the former phase quarks and gluons act as a degree of freedom whereas in the latter, mesons and baryons play this role. In QGP phase the chiral symmetry is followed ($m_q \sim 0$) but in hadronic phase it is broken explicitly ($m_q \neq 0$) and spontaneously ($\langle \bar q q \rangle \neq 0$)   \cite{Rapp2010,Vogt2007}.   Furthermore, in the hadronic ensemble, the thermodynamics quantities namely nuclear density (number density of nucleons), isospin asymmetry (number of neutrons vs the number of protons), and temperature also play a crucial role to modify the in-medium properties of the mesons and baryons \cite{Papazoglou1999,Tolos2020,Vogt2007}. 
	The operation of future experimental facilities  such as CBM and PANDA at GSI, Germany, NICA at Dubna, Russia and, J-PARC at Japan may lead to considerable  progress in the understanding of meson-baryons interactions   \cite{Vogt2007,Rapp2010}.
	
	On the theoretical side,  several potential models have been theorized to study the physics of the non-perturbative regime.  Some of these are: Nambu-Jona-Lasinio (NJL) model \cite{Nambu1961},  the Polyakov loop extended NJL  (PNJL) model \cite{Fukushima2004,Kashiwa2008,Ghosh2015}, chiral perturbation theory (ChPT) \cite{Zhong2006,Jenkins1991}, coupled channel approach \cite{Tolos2020,Tolos2004,Tolos2006,Tolos2008,Hofmann2005}, chiral $SU(3)$ model \cite{Papazoglou1999,Mishra2004,Mishra2006,Kumar2010,Kumar2011,Kumar2019a,Kumar2020b,Mishra2004a,Mishra2009,Kumar2010,
		Kumar2019,Kumar2020}, Quark-Meson Coupling (QMC) model \cite{Guichon1988,Hong2001,Tsushima1999,Sibirtsev1999,Saito1994,Panda1997},  Polyakov Quark Meson (PQM) model \cite{Chatterjee2012,Schaefer2010},    and QCD sum rules  \cite{Reinders1981,Hayashigaki2000,
		Hilger2009,Reinders1985,Klingl1997,Klingl1999}, etc. Various effective models are formulated keeping in view the fundamental QCD properties such as broken scale invariance and spontaneous and explicit breaking of the chiral symmetry.

	
	For the first time,
	Haider and Liu   anticipated that the $\eta N$ interactions
	are attractive and suggested the possibility of $\eta$-meson bound states \cite{Haider1986,Liu1986}. The negative mass-shift/optical potential of $\eta$-meson has attracted researchers to study the possibilities of $\eta$-mesic nuclei formation \cite{Jenkins1991,Zhong2006,Waas1997}. At nuclear saturation density, the optical potential of -20 MeV was anticipated in the chiral coupled channel approach, considering  leading order terms \cite{Waas1997}. Using  same coupled channel model,  Chiang $et. al.$ obtained optical potential $U_{\eta}$ = -34 MeV in the normal nuclear matter, assuming the $\eta N$ interactions  dominated by  $N^*$(1535) excitation  \cite{Chiang1991} and anticipated that  the attractive potential can produce $\eta$-meson bound state with light and heavy nucleus. Using the QMC  model, authors of Ref. \cite{Tsushima1998} obtained  optical potential -60 MeV at $\rho_{N}=\rho_0$. The chiral unitary approach was also implied to evaluate the $\eta$ potential and it was observed to be -54 MeV \cite{Inoue2002}.  A more deep optical potential of -72 MeV was  observed in Ref. \cite{Wang2010}. In this article the possibility of a bound state with $\eta$-meson was also explored. 
	
	In  Ref. \cite{Zhong2006}, using $\eta N$ Lagrangian off-shell terms, at normal
	nuclear density, the  in-medium mass of $\eta$-meson was found to be  (0.84 $\pm$ 0.015)$m_{\eta}$
	and the corresponding  optical potential was observed as  -(83 $\pm$ 5) MeV.   Furthermore, using the relativistic mean-field theory, Song $et$ $al.$ observed the optical potential by varying the scattering length \cite{Song2008}. Clearly, the values of $\eta$ optical potential predicted in various studies varies over large range, $i.e.,$ -20 MeV to -85 MeV and hence, have considerable model dependence.  In addition to theoretical attempts, there are experimental studies to explore the properties of $\eta$ mesons \cite{Peng1987,Berg1994,Chiavassa1998,Martinez1999,Averbeck2003,Agakishiev2013}. For example, for different $\eta$ hadron interactions, the $\eta$-meson production has been studied  in Refs. \cite{Peng1987,Martinez1999,Agakishiev2013} and the transverse momentum spectra of $\eta$-meson is measured in HICs near the free $N$-$N$ production threshold \cite{Agakishiev2013}.

	%

	In the current investigation,  we present the in-medium mass and optical potential of the $\eta$-meson in  hot and dense asymmetric nuclear matter using chiral SU(3) model. By incorporating the  medium induced  nucleon scalar densities,  we calculate the in-medium mass-shift of $\eta$-meson using  the  $\eta N$ effective Lagrangian of chiral SU(3) model. Furthermore,	as discussed earlier, the in-medium mass and optical potential of $\eta$-meson have been studied using the unitary approach of ChPT and relativistic mean-field model \cite{Zhong2006,Song2008}. Following this work, as a second part of the current investigation the effective mass of $\eta$-meson is also evaluated using the chiral $\eta N$  Lagrangian of chiral perturbation theory \cite{Zhong2006}. In this approach, the nucleon scalar densities are calculated from chiral SU(3) model and  plugged in the dispersion relation of $\eta N$ interactions derived from ChPT Lagrangian.

	The chiral $SU(3)$ model is extensively used   to explore the in-medium  properties of the mesons and baryons in the hot and dense hadronic matter \cite{Kumar2010,Zschiesche2004,Mishra2004}. For example, the model was used to study the in-medium mass and optical potential of   kaons, antikaons and phi mesons in the nuclear and hyperonic matter \cite{Mishra2004,Kumar2011,Kumar2020b}. Furthermore, in the nuclear and hadronic matter the  in-medium mass of spin 0, spin 1 $D$ mesons and quarkonia were calculated using the conjunction of chiral SU(3) model and QCD sum rules with \cite{Kumar2020,Kumar2020a,Kumar2019,Kumar2019a} and without taking the effect of magnetic field \cite{Kumar2014,Chhabra2017,Chhabra2017a,Chhabra2018,Kumar2010}.  The model was extended to $SU(4)$ and $SU(5)$ sector to evaluate the  medium induced  properties of heavy mesons such as $D$ and $B$  \cite{Mishra2004a,Mishra2009,Kumar2011}. On the other hand, the chiral perturbation theory is also a successful theoretical  framework to study the baryon-meson interactions. The in-medium properties of $K$ meson were first studied by Kaplan and Nelson using chiral perturbation theory (ChPT) \cite{Kaplan1986}.  The same theory was applied to study the $\eta$-nucleon interactions via adding leading order  terms in the model Lagrangian \cite{Jenkins1991}. The heavy baryon chiral perturbation theory was also applied to study the kaon condensation which is an imperative property to study the neutron star matter \cite{Brown1994,Lee1995,Kaiser1995}. The ChPT theory was also improved by the introduction of next-to-leading  order terms in the chiral effective Lagrangian. By including these off-shell terms, Zhong $et. al.$ anticipated appreciable decrease in the in-medium mass of $\eta$-meson which is  favorable for the formation of $\eta$-mesic nuclei \cite{Zhong2006}.

	The layout of the present paper is as follows: 
	In the next section, we will give brief explanation of the formalism used in the present work. In  section \ref{subsec2.1.1}, we will derive the $\eta N$ interactions in the chiral SU(3) model whereas, in section \ref{subsec2.1.2}, $\eta N$ methodology will be given in the  unified approach of chiral perturbation theory and chiral model. In section \ref{sec:3}, we will discuss the in-medium effects on the mass of $\eta$-meson, and finally in section \ref{sec:4}, we will present the summary.

	\section{ FORMALISM }

	\subsection{IN-MEDIUM SCALAR FIELDS IN THE CHIRAL SU(3) MODEL}
	\label{subsec2.1}

	The Lagrangian density of the chiral SU(3) model is written as
	\be
	{\cal L}_{\text{chiral}} = {\cal L}_{kin} + \sum_{ M =S,V}{\cal L}_{NM}
	+ {\cal L}_{vec} + {\cal L}_0 + {\cal L}_{SB}.
	\label{genlag} \ee 
	
	The model preserves the fundamental QCD properties such as the  broken scale invariance and non-linear realization of the chiral symmetry \cite{Weinberg1968,Coleman1969,Zschiesche1997,Bardeen1969,Kumar2020,Papazoglou1999,Kumar2019}. It is successfully used  to explain the nuclear matter, finite nuclei, neutron star, and  hypernuclei \cite{Weinberg1968,Coleman1969,Zschiesche1997,Bardeen1969,
		Kumar2020,Papazoglou1999,Kumar2019}. In this model,  the nucleons and baryons interact by the exchange of the vector fields $\omega$ and $\rho$  along with the scalar fields $\sigma$, $\zeta $ and $\delta$  in the nuclear medium. The vector fields give short-range repulsion or attraction which depends on the type of meson-nucleon interaction whereas the scalar fields give attractive contributions to the medium \cite{Kumar2020b}. The $\sigma$ field is a  non-strange scalar-isoscalar field  which represents the scalar mesons $\sigma$ ($u\bar{d}$) whereas the $\zeta$ field is a strange  scalar-isoscalar field which represent the scalar meson ($s\bar{s}$) \cite{Zakout2000}. Moreover,  the scalar-isovector field $\delta$ $\sim (\bar u  u-\bar d d$)  is incorporated in the present model to study the effect of  the isospin asymmetric matter. Further, the  glueball field, $\chi$  is a hypothetical gluon field that contains  gluon particles and  is   introduced in the chiral models to incorporate the scale invariance property of QCD \cite{Papazoglou1999,Pwang2001}. We have used mean-field approximation to simplify the model by neglecting the effect of quantum and thermal  fluctuations near phase transitions  \cite{Kumar2020,Reddy2018}.

	In Eq.(\ref{genlag}), the  ${\cal L}_{kin}$ term   describes the the kinetic energy term and the second term ${\cal L}_{NM}$ given by 
	\begin{eqnarray}
		{\cal L}_{NM} = - \sum_{i} \bar {\psi_i} 
		\left[ m_{i}^{*} + g_{\omega i} \gamma_{0} \omega 
		+ g_{\rho i} \gamma_{0} \rho \right] \psi_{i},
		\label{NM}
	\end{eqnarray}
	
	defines the nucleon-meson interactions	with in-medium nucleon mass  as
		\begin{eqnarray}
	  m_{i}^{*} = -(g_{\sigma i}\sigma + g_{\zeta i}\zeta + g_{\delta i}\tau_3 \delta),
		\label{massn}
	\end{eqnarray}
 where $\tau_3$ denotes the 3$^{rd}$ component of isospin and $g_{\sigma i}$, $g_{\zeta i}$ and $g_{\delta i}$ are  the coupling constants of  $\sigma$,  $\zeta$ and  field $\delta$ with  nucleons ($i$=$p,n$), respectively.  The next term $ {\cal L}_{vec}$ is given by
	
	\begin{eqnarray}
		{\cal L} _{vec} & = & \frac {1}{2} \left( m_{\omega}^{2} \omega^{2} 
		+ m_{\rho}^{2} \rho^{2} \right) 
		\frac {\chi^{2}}{\chi_{0}^{2}}
		+  g_4 (\omega ^4 +6\omega^2 \rho^2+\rho^4),
		\label{vec}
	\end{eqnarray}
	
	reproduces the mass of vector mesons through self-interactions. The ${\cal L}_{0}$ defines  the spontaneous chiral symmetry breaking by the equation
	
	\begin{eqnarray}
		{\cal L} _{0} & = & -\frac{1}{2} k_{0}\chi^{2} \left( \sigma^{2} + \zeta^{2} 
		+ \delta^{2} \right) + k_{1} \left( \sigma^{2} + \zeta^{2} + \delta^{2} 
		\right)^{2} \nonumber\\
		&+& k_{2} \left( \frac {\sigma^{4}}{2} + \frac {\delta^{4}}{2} + 3 \sigma^{2} 
		\delta^{2} + \zeta^{4} \right) 
		+ k_{3}\chi\left( \sigma^{2} - \delta^{2} \right)\zeta \nonumber\\
		&-& k_{4} \chi^{4} 
		-  \frac {1}{4} \chi^{4} {\rm {ln}} 
		\frac{\chi^{4}}{\chi_{0}^{4}}
		+ \frac {d}{3} \chi^{4} {\rm {ln}} \Bigg (\bigg( \frac {\left( \sigma^{2} 
			- \delta^{2}\right) \zeta }{\sigma_{0}^{2} \zeta_{0}} \bigg) 
		\bigg (\frac {\chi}{\chi_0}\bigg)^3 \Bigg ).
		\label{lagscal}
	\end{eqnarray}

	In this equation, the  $\sigma_0$, $\zeta_0$, $\delta_0$ and $\chi_0$ denote the vacuum values of  $\sigma$, $\zeta$, $\delta$ and $\chi$ scalar fields, respectively.  Also, the parameter $d$=0.064  along with  $k_i(i=1$ to $4)$ and other medium parameters are fitted to  regenerate the vacuum values of scalar and vector fields, $\eta$, $\eta'$ 
	mesons  and  the   nucleon mass  \cite{Papazoglou1999,Kumar2010,Kumar2019}. In \cref{ccc}, we have tabulated the values  of various parameters. 
	Furthermore, the  last term ${\cal L}_{SB} $ in Eq.(\ref{genlag})  describes the explicit chiral symmetry breaking property and is written as

	\begin{eqnarray}
		{\cal L} _{SB} =  -\left( \frac {\chi}{\chi_{0}}\right)^{2} 
		\left[ m_{\pi}^{2} 
		f_{\pi} \sigma
		+ \big( \sqrt {2} m_{K}^{2}f_{K} - \frac {1}{\sqrt {2}} 
		m_{\pi}^{2} f_{\pi} \big) \zeta \right].
		\label{lsb}
	\end{eqnarray} 
	
	In the above equation,   $m_\pi$, $m_K$, $f_\pi$, and $f_K$  symbolize  the masses and decay constants of pions and kaons, respectively.  
	%

	The non-linear coupled equations of motion  of the scalar and vector fields are deduced by solving   the total Lagrangian (Eq.(\ref{genlag})) using the Euler-Lagrange equations \cite{Kumar2019,Kumar2019a} and are given as

	\begin{eqnarray}
		k_{0}\chi^{2}\sigma-4k_{1}\left( \sigma^{2}+\zeta^{2}
		+\delta^{2}\right)\sigma-2k_{2}\left( \sigma^{3}+3\sigma\delta^{2}\right)
		-2k_{3}\chi\sigma\zeta \nonumber\\
		-\frac{d}{3} \chi^{4} \bigg (\frac{2\sigma}{\sigma^{2}-\delta^{2}}\bigg )
		+\left( \frac{\chi}{\chi_{0}}\right) ^{2}m_{\pi}^{2}f_{\pi}
		=\sum g_{\sigma i}\rho_{i}^{s} ,
		\label{sigma}
	\end{eqnarray}
	\begin{eqnarray}
		k_{0}\chi^{2}\zeta-4k_{1}\left( \sigma^{2}+\zeta^{2}+\delta^{2}\right)
		\zeta-4k_{2}\zeta^{3}-k_{3}\chi\left( \sigma^{2}-\delta^{2}\right)\nonumber\\
		-\frac{d}{3}\frac{\chi^{4}}{\zeta}+\left(\frac{\chi}{\chi_{0}} \right)
		^{2}\left[ \sqrt{2}m_{K}^{2}f_{K}-\frac{1}{\sqrt{2}} m_{\pi}^{2}f_{\pi}\right]
		=\sum g_{\zeta i}\rho_{i}^{s} ,
		\label{zeta}
	\end{eqnarray}
	\begin{eqnarray}
		k_{0}\chi^{2}\delta-4k_{1}\left( \sigma^{2}+\zeta^{2}+\delta^{2}\right)
		\delta-2k_{2}\left( \delta^{3}+3\sigma^{2}\delta\right) +2k_{3}\chi\delta
		\zeta \nonumber\\
		+   \frac{2}{3} d \chi^4 \left( \frac{\delta}{\sigma^{2}-\delta^{2}}\right)
		=\sum g_{\delta i}\tau_3\rho_{i}^{s}  ,
		\label{delta}
	\end{eqnarray}

	\begin{eqnarray}
		\left (\frac{\chi}{\chi_{0}}\right) ^{2}m_{\omega}^{2}\omega+g_{4}\left(4{\omega}^{3}+12{\rho}^2{\omega}\right) =\sum g_{\omega i}\rho_{i}^{v}  ,
		\label{omega}
	\end{eqnarray}

	\begin{eqnarray}
		\left (\frac{\chi}{\chi_{0}}\right) ^{2}m_{\rho}^{2}\rho+g_{4}\left(4{\rho}^{3}+12{\omega}^2{\rho}\right)=\sum g_{\rho i}\tau_3\rho_{i}^{v}  ,
		\label{rho}
	\end{eqnarray}
	
	and
	
	\begin{eqnarray}
		k_{0}\chi \left( \sigma^{2}+\zeta^{2}+\delta^{2}\right)-k_{3}
		\left( \sigma^{2}-\delta^{2}\right)\zeta + \chi^{3}\left[1
		+{\rm {ln}}\left( \frac{\chi^{4}}{\chi_{0}^{4}}\right)  \right]
		+(4k_{4}-d)\chi^{3}
		\nonumber\\
		-\frac{4}{3} d \chi^{3} {\rm {ln}} \Bigg ( \bigg (\frac{\left( \sigma^{2}
			-\delta^{2}\right) \zeta}{\sigma_{0}^{2}\zeta_{0}} \bigg )
		\bigg (\frac{\chi}{\chi_0}\bigg)^3 \Bigg )+
		\frac{2\chi}{\chi_{0}^{2}}\left[ m_{\pi}^{2}
		f_{\pi}\sigma +\left(\sqrt{2}m_{K}^{2}f_{K}-\frac{1}{\sqrt{2}}
		m_{\pi}^{2}f_{\pi} \right) \zeta\right] \nonumber\\
		-\frac{\chi}{{{\chi_0}^2}}(m_{\omega}^{2} \omega^2+m_{\rho}^{2}\rho^2)  = 0 ,
		\label{chi}
	\end{eqnarray}

	respectively.
	
	In  above equations, the $\rho^{s}_{i}$ and   $\rho^{v}_{i}$   denote  the  scalar and vector densities of $i^{th}$ nucleons ($i=n,p$) \cite{Kumar2019,Papazoglou1999} and  are given as
	
	\begin{eqnarray}
		\rho_{i}^{v} = \gamma_{i}\int\frac{d^{3}k}{(2\pi)^{3}}  
		\Bigg(\frac{1}{1+\exp\left[\beta(E^{\ast}_i(k) 
			-\mu^{*}_{i}) \right]}-\frac{1}{1+\exp\left[\beta(E^{\ast}_i(k)
			+\mu^{*}_{i}) \right]}
		\Bigg),
		\label{rhov0}
	\end{eqnarray}
	
	and
	
	\begin{eqnarray}
		\rho_{i}^{s} = \gamma_{i}\int\frac{d^{3}k}{(2\pi)^{3}} 
		\frac{m_{i}^{*}}{E^{\ast}_i(k)} \Bigg(\frac{1}{1+\exp\left[\beta(E^{\ast}_i(k) 
			-\mu^{*}_{i}) \right]}+\frac{1}{1+\exp\left[\beta(E^{\ast}_i(k)
			+\mu^{*}_{i}) \right]}
		\Bigg),
		\label{rhos0}
	\end{eqnarray}
	respectively, where $\beta = \frac{1}{kT}$, $E^{\ast}_i(k)=\sqrt{k^2+{m^*_i}^2}$, $ \mu^{*}_{i}=\mu_{i}-g_{\omega i}\omega-g_{\rho i}\tau_{3}\rho$  and $\gamma_i$ is the degeneracy factor. Moreover, the  isospin effect on the scalar and vector density is measured by the definition, $I = -\frac{\Sigma_i \tau_{3i} \rho^{v}_{i}}{2\rho_{N}}$. In the next section, we calculate the medium-modified  mass of  $\eta$ mesons 
	in hot asymmetric  nuclear matter. The medium modified $\eta$ meson mass is evaluated from the dispersion relation which is obtained from the $\eta N$ equation of motion.
	
	\begin{table}
		\begin{tabular}{|c|c|c|c|c|}

			\hline
			$k_0$ & $k_1$ & $k_2$ & $k_3$ & $k_4$  \\ 
			\hline 
			2.53 & 1.35 & -4.77 & -2.77 & -0.218  \\ 
			\hline

			\hline 
			$\sigma_0$ (MeV)& $\zeta_0$ (MeV) & $\chi_0$ (MeV)  & $d$ & $\rho_0$ ($\text{fm}^{-3}$)  \\ 
			\hline 
			-93.29 & -106.8 & 409.8 & 0.064 & 0.15  \\

			\hline
			$g_{\sigma N}$  & $g_{\zeta N }$  &  $g_{\delta N }$  &
			$g_{\omega N}$ & $g_{\rho N}$ \\

			\hline 
			10.56 & -0.46 & 2.48 & 13.35 & 5.48  \\

			\hline 
			$m_\pi $ (MeV) &$ m_K$ (MeV)&$ f_\pi$ (MeV)  & $f_K$ (MeV) & $g_4$ \\ 
			\hline 
			139 & 498 & 93.29 & 122.14 & 79.91  \\ 
			\hline
			
			$m_{\sigma}$ (MeV) & $m_{\zeta}$ (MeV)  &  $m_{\delta}$  (MeV) &
			$M_N$ (MeV) &  $m_{\eta}$ (MeV) \\
			\hline
			
			466.5 & 1024.5 & 899.5 &  939&574.374 \\ 
			\hline

		\end{tabular}
		\caption{Different constants used in the present work \cite{Papazoglou1999}.} \label{ccc}
	\end{table} 
	
	\subsubsection{$\eta$N INTERACTIONS IN THE CHIRAL SU(3) MODEL}
	\label{subsec2.1.1}

\
In the chiral SU(3) model, the $\eta N$ interaction Lagrangian density can be written as
	\begin{eqnarray} \label{etaN}
		\mathcal{L_{\eta }}  &=&
		\left( \frac{1}{2}-\frac{\sigma ^\prime + 4 \zeta ^\prime (2 f_K-f_\pi) }{\sqrt{2}f^2} \right) \partial^{\mu}\eta\partial_{\mu}\eta \nonumber\\
		&-&\frac{1}{2}\left(
		m_{\eta}^2
		-\frac{(\sqrt{2}\sigma ^\prime -4 \zeta ^\prime )m^2_\pi f_\pi + 8 \zeta ^\prime m^2_K f_K}{\sqrt{2} f^2}
		\right) \eta^2\nonumber\\
		&&+\frac{d'}{f^2} \left( \frac{\rho^s_p+\rho^s_n}{4}  \right) \partial^{\mu}\eta\partial_{\mu}\eta,
	\end{eqnarray}

		The above chiral $\eta N$ Lagrangian  consists of three terms.
		
		\begin{itemize}
		    \item 	First Range Term:
		    
		The first term in the chiral Lagrangian describes  the first range term \cite{Kumar2011,Papazoglou1999}  and is obtained from
	\begin{equation}
		\label{pikin}
		{\mathcal L}_{{\mathrm{1st range term}}} =  Tr (u_{\mu} X u^{\mu}X +X u_{\mu} u^{\mu} X) . 
	\end{equation}
	In the above equation,  $u_{\mu} =-\frac{i}{2} \left[u^{\dagger}(\partial_{\mu}u) 
	-u (\partial_{\mu}u^\dagger) \right]$ and $
	u$=$ \text{exp}\left[ \frac{i}{\sqrt{2}\sigma_{0}}P\gamma_{5}\right]$, which is expanded  up to second order. Here,  symbols $X$ and $P$ represent the scalar and pseudoscalar   meson matrices  \cite{Papazoglou1999}, respectively and are explicitly given by  Eqs. (\ref{smat}) and (\ref{psmat}) in the \cref{appendix}. Furthermore, the vacuum values of $\sigma$ and $\zeta$ fields are deduced in terms  of pions and kaons  decay constant by solving the axial current of  pions and kaons \cite{Papazoglou1999} through relation
	
	\be
	\label{zeta0}
	\sigma_0 = -f_{\pi} \qquad \zeta_0 = -\frac{1}{\sqrt{2}}(2 f_K - f_{\pi}).
	\ee

	Moreover, in the first term of $\eta N$ Lagrangian   $\sigma'(=\sigma-\sigma_0)$,
	$\zeta'(=\zeta-\zeta_0)$ and  $\delta'(=\delta-\delta_0)$ define the  digression of the expectation values of fields
	from their vacuum expectations. Also, $f$=$\sqrt{f_\pi^2+2(2 f_K - f_\pi)^2}$ and  $d'$=$3d_1+d_2$ are the constant parameters.

	\item 	Mass Term:
	
	Further, the mass term of the chiral model gives the second term of $\eta N$ Lagrangian  and is given by
	
	\begin{equation}
		\label{esb-gl}
		{\cal L}_{SB}  =  
		-\frac{1}{2} \Tr A_p \left(uXu+u^{\dagger}Xu^{\dagger}\right),
	\end{equation}
	
	where $A_p$ is a diagonal matrix given in the Eq.(\ref{apmat}). The vacuum mass of $\eta$ meson, $m_{\eta}$ , is extracted from the above term and is given by the relation

	\begin{equation}
		m_{\eta}=\frac{1}{f}\sqrt{\left(3 m_\pi^2  f_K m_K^2+\frac{8 f_K^2 m_K^2}{f_\pi^2} -\frac{4f_K  m_\pi^2}{f_\pi} \right)}.
	\end{equation}
	
	Substituting the values of various constants in above $m_{\eta}$ turns out to be 574.374 MeV which is with an accuracy of 4.9 $\%$  of physical mass $i.e.$ 547.862 MeV \cite{PDG2020}. The vacuum mass of $\eta$-meson has  model dependency \cite{Burakovsky1997} but here in the present work, we are more concerned in the $\eta$ in-medium mass-shift which is nearly same for both the masses. In Ref. \cite{Burakovsky1997},  using Gell-Mann Okubo mass formula under octet approximation in the SU(4) meson multiplets, authors calculated the vacuum mass of $\eta$-meson to be 567 MeV which  is with an accuracy of 3.6 $\%$.

	\item 	 $d'$  Term:
	
	The third term ($i.e.$ $d'$ term) in the $\eta N$ Lagrangian originates from the baryon-meson interaction Lagrangian densities   \cite{Mishra2004a,Mishra2006}  
	\begin{equation}
		{\cal L }_{d_1}^{BM} =\frac {d_1}{2} Tr (u_\mu u ^\mu)Tr( \bar B B),
	\end{equation}
	and
	\begin{equation}
		{\cal L }_{d_2}^{BM} =d_2 Tr (\bar B u_\mu u ^\mu B).
	\end{equation}
	In above,  $B$ denotes the baryon matrix (see Eq.(\ref{bmat})). 
	
	\end{itemize}

	It should be noted that in case of $\eta N$ interactions of Eq. (\ref{etaN}), the terms corresponding to vectorial Weinberg-Tomozawa term vanishes.
	On the the other hand, the Weinberg-Tomozawa term plays a crucial role in the determination of $K (\bar K)$ and $D (\bar D)$ in-medium mass \cite{Kumar2011,Mishra2009}.

	Using the $\eta N$ Lagrangian in the Euler-Lagrange equation for $\eta$ meson,
	the equation of motion  is evaluated as
	\begin{eqnarray} 
		&& \partial^{\mu}\partial_{\mu} \eta -\left(
		m_{\eta}^2-\frac{(\sqrt{2}\sigma ^\prime -4 \zeta ^\prime )m^2_\pi f_\pi + 8 \zeta ^\prime m^2_K f_K}{\sqrt{2} f^2}
		\right)\eta  \nonumber\\
		&&+\frac{2d'}{f^2} \left( \frac{\rho^s_p+\rho^s_n}{4} -\frac{\sigma ^\prime + 4 \zeta ^\prime (2 f_K-f_\pi) }{\sqrt{2}} \right) \partial^{\mu} \partial_{\mu}  \eta=0.
	\end{eqnarray}
	
Performing the Fourier transformation  on the above equation,  the dispersion relation for $\eta$ meson turns out to be
	\begin{equation}
		-\omega^2+ { \textbf{k}}^2 + m_\eta^2 -\Pi^*(\omega, | \textbf{k}|)=0.
		\label{drk}
	\end{equation}
	
	In the above equation,  $\Pi^*$ denotes the effective self-energy of  $\eta$ meson, explicitly given as
	\begin{eqnarray}
		\Pi^* (\omega, | \textbf{k}|) &= & -
		\frac{(\sqrt{2}\sigma ^\prime -4 \zeta ^\prime )m^2_\pi f_\pi + 8 \zeta ^\prime m^2_K f_K}{\sqrt{2} f^2}
		+\frac{2d'}{f^2} \left( \frac{\rho^s_p+\rho^s_n}{4} \right)
		(\omega ^2 - {\vec k}^2) \nonumber\\
		&-& \frac{2}{f^2} \left( \frac{\sigma ^\prime + 4 \zeta ^\prime (2 f_K-f_\pi) }{\sqrt{2}} \right)
		(\omega ^2 - {\textbf{ k}}^2).
		\label{sen}
	\end{eqnarray}

	The unknown parameter, $d'$ is approximated from the experimental values of scattering length, $a^{\eta N}$ \cite{Zhong2006}. In the chiral model, the expression of scattering length  derived from the scattering amplitude is given by

	\begin{eqnarray}
		a^{\eta N} &=& \frac{1}{4 \pi \left (1+\frac{m_\eta}{M_N}\right )} \Big [ \Big( \frac{d'}{\sqrt{2}}-\frac{g_{\sigma N}}{m^2_\sigma}+\frac{4 (2f_K-f_\pi) g_{\zeta N}}{m^2_\zeta} \Big) \frac {m_\eta ^2} {\sqrt{2}f^2} \nonumber \\
		&+& \left( \frac{\sqrt{2} g_{\sigma N}}{m^2_\sigma}-\frac{4 g_{\zeta N}}{m^2_\zeta} \right )\frac {m^2_\pi f_\pi} {2\sqrt{2}f^2}+\tau _3 \frac{2\sqrt{2} g_{\delta N}}{m^2_\delta} \frac {m^2_K f_K} {f^2}  \Big ].
		\label{sl}
	\end{eqnarray}
	Rearranging the above for  $d'$ gives
	\begin{eqnarray}
		d' &=& \frac{ f^2}{2 \pi \left (1+\frac{m_\eta}{M_N}\right )} \frac{ a^{\eta N}}{m^2_\eta} +\frac{\sqrt{2}g_{\sigma N}}{m^2_\sigma}-\frac{4 \sqrt{2} (2f_K-f_\pi) g_{\zeta N}}{m^2_\zeta}  \nonumber \\
		&-& \left( \frac{\sqrt{2} g_{\sigma N}}{m^2_\sigma}-\frac{4 g_{\zeta N}}{m^2_\zeta} \right )\frac {m^2_\pi f_\pi} {\sqrt{2} m^2_\eta}-\tau _3 \frac{4\sqrt{2} g_{\delta N}m^2_K f_K}{m^2_\delta m^2_\eta}.
		\label{dp}
	\end{eqnarray}

	Using the condition, $m_{\eta}^*=\omega(| \textbf{k}|$=0) in Eq. (\ref{drk}), we obtain the effective mass of $\eta$ meson in the nuclear  medium. Further, the momentum dependent optical potentials are defined through the relation \cite{Mishra2008,Mishra2009} 
	\begin{equation}
		U^*_{\eta}(\omega,\textbf{ k}) = \omega (\textbf{k}) -\sqrt {\textbf{k}^2 + m^{^2}_{\eta}}.
	\end{equation}
At zero momentum, the above equation gives
	\begin{equation}
		U^*_{\eta} =\Delta m_\eta^*={m_{\eta}^*}-m_{\eta}.
		\label{opc}
	\end{equation}

	\subsubsection{\label{subsec2.1.2} UNIFICATION OF CHIRAL PERTURBATION THEORY (ChPT) AND CHIRAL MODEL}

	In this section, we discuss the unified approach of ChPT and chiral model to compute the in-medium mass of $\eta$ mesons.  The ChPT comprises the underlying chiral symmetry property of QCD and use an effective field theory approach \cite{Zhong2006}. The same theory along with Relativistic mean-field model   has been used  to deduce the eta-nucleon interactions in the symmetric nuclear matter \cite{Zhong2006,Song2008}. The  Lagrangian density defining the meson-baryons interactions in this theory is given by
	
	\begin{eqnarray}\label{LL}
		{\mathcal{L}_{\text{ChPT}} }={\mathcal{L}_{P} }+{\mathcal{L}_{P
				B} },
	\end{eqnarray}
	with $P$ representing the pseudoscalar meson  multiplet (see Eq.(\ref{psmat})).  Up to second chiral order, the $\mathcal{L}_{P}$ term  is defined as \cite{Zhong2006,Kaplan1986}
	
	\begin{eqnarray}
		{\mathcal{L}_{P} }&=&\frac{1}{4}f_\pi^{2}\textrm{Tr}
		\partial^{\mu}\Sigma\partial_{\mu}\Sigma^{\dagger}
		+\frac{1}{2}f_\pi^2 B_0
		\left\{\mbox{Tr} M_{q}(\Sigma-1)+\mathrm{h.c.}\right\},
	\end{eqnarray}

	where $\Sigma=\xi^2=\exp{(i\sqrt{2}P/f_\pi)}$ and $M_{q}=\mbox{diag}\{m_{q}, m_{q}, m_{s}\}$
	is the current quark mass matrix.  The 
	Lagrangian term, $\mathcal{L}_{P B}$=$\mathcal{L}^L_{P B}$+$\mathcal{L}^{NL}_{P B}$ describes the leading and next to leading order contributions \cite{Kaplan1986}.  Jenkins and Manohar developed the next to leading order terms  using 
	heavy baryon chiral theory \cite{Jenkins1991}. In this Lagrangian, the loop contributions are not considered as the higher-order corrections get suppressed for the  small
	momentum scale, $Q^{2}$ \cite{Zhong2006}. The  different nuclear properties are studied successfully  using $\mathcal{L}^{NL}_{P B}$ \cite{Park1993}.

	The $\eta N$ Lagrangian is obtained by expanding the Eq.(\ref{LL}) up to the second order of multiplet $P$ \cite{Zhong2006}
	\begin{eqnarray} \label{teffL}
		\mathcal{L_{\eta N }}  &=&
		\frac{1}{2}\partial^{\mu}\eta\partial_{\mu}\eta
		-\frac{1}{2}\left(
		m{^\prime}_{\eta}^2
		-\frac{\Sigma_{\eta
				\mathrm{N}}}{f_\pi^2}\bar{\Psi}_{\mathrm{N}}\Psi_{\mathrm{N}}
		\right) \eta^2+\frac{1}{2}\frac{\kappa}{f_\pi^2}\bar{\Psi}_{\mathrm{N}}\Psi_{\mathrm{N}}\partial^{\mu}\eta\partial_{\mu}\eta.
	\end{eqnarray}
	Here,  $m{^\prime}_{\eta}=\sqrt{\frac{2}{3}B_0 (m_q+2m_s)}$ denotes  the vacuum mass  of $\eta$ meson calculated in chiral perturbation theory. In the mass expression,  $B_0$ symbolize the relation with the order
	parameter of spontaneously broken chiral symmetry and $m_{q(s)}$ denote the mass of light (strange) quarks \cite{Burakovsky1997}. For consistency with the chiral SU(3) model, we have used the same value of  $\eta$ meson vacuum mass $i.e.$ $m{^\prime}_{\eta}$=$m_{\eta}$= 574.374 MeV in the further calculations of ChPT. The $\eta N$ sigma term $\Sigma_{\eta\mathrm{N}}$, obtained from \enquote{$a_i$} terms of the next-to-leading order chiral Lagrangian density is given as \cite{Zhong2006}
	\begin{eqnarray} \label{sigmaexp}
		\Sigma_{\eta\mathrm{N}}
		=-\frac{2}{3}[a_{1}m_{q}+4a_{2}m_{s}+2a_{3}(m_{q}+2m_{s})].
	\end{eqnarray}
	
	The $\Sigma_{\eta\mathrm{N}}$ value is estimated to be 280 $\pm$ 130 MeV from the different empirical observations of  $\Sigma_{\mathrm{KN}}$ term having value 380 MeV $\pm$ 100 MeV \cite{Lyubovitskij2001,Dong1996,Hatsuda1994,
		Brown1994,Georgi1984,Politzer1991,Lee1995,Zhong2006}.
	
	Also, the parameter $\kappa$ in the last term of the Eq.(\ref{teffL}) comprises the contributions from the  \enquote{off-shell} $d_i$ terms of the next to Leading order Lagrangian \cite{Zhong2006}. In the present work, we determined $\kappa$  using the expression of $\eta N$ scattering length, 
	$a^{\eta\mathrm{N}}$, calculated from the ChPT matrix amplitude  (on-shell constraints) \cite{Zhong2006}
	
	\begin{eqnarray}
		a^{\eta\mathrm{N}} =\frac{1}{4\pi f_\pi^2(1+m_{\eta}/M_{\mathrm{N}})}
		\left(\Sigma_{\eta\mathrm{N}}
		+ \kappa m_{\eta}^2\right),
	\end{eqnarray}
	and by re-arranging for $\kappa$
	it becomes 
	
	\begin{eqnarray} \label{akap}
		\kappa =4\pi f_\pi^2
		\left(
		\frac{1}{m_{\eta}^2}+\frac{1}{m_{\eta}M_{\mathrm{N}}}
		\right)
		a^{\eta\mathrm{N}} -\frac{\Sigma_{\eta\mathrm{N}}}{m_{\eta}^2}.
		\label{kappa}
	\end{eqnarray}

	We have taken the experimentally determined $a^{\eta\mathrm{N}}$ values $i.e.$ 0.91 $\sim$ 1.14 fm in the present investigation \cite{Green2005,Renard2002,Arndt2005,Green1999,Zhong2006}. Furthermore, the $\eta N$ equation of motion has been derived using the interaction Lagrangian (Eq.(\ref{teffL})) in the Euler Lagrange equation of motion:

	\begin{eqnarray}
		\left(
		\partial_{\mu}\partial^{\mu}
		+m_{\eta}^2
		-\frac{\Sigma_{\eta N}}{2f_\pi^2}\langle
		\bar{\Psi}_{\mathrm{N}}\Psi_{\mathrm{N}} \rangle
		+\frac{\kappa}{2f_\pi^2}\langle
		\bar{\Psi}_{\mathrm{N}}\Psi_{\mathrm{N}} \rangle\partial_{\mu}\partial^{\mu}
		\right)
		\eta
		= 0.
		\label{eqm}
	\end{eqnarray}

	In above, $ \langle
	\bar{\Psi}_{\mathrm{N}}\Psi_{\mathrm{N}}\rangle \equiv \rho^s_{N}$=$\left(\rho^s_{p}+\rho^s_{n} \right)$  defines the  in-medium scalar
	density of nucleons calculated within the mean-field chiral SU(3) model (see Eqs.(\ref{rhov0}) and (\ref{rhos0})). The Fourier transformation  of Eq.(\ref{eqm}) gives
	\begin{eqnarray} \label{decom}
		-\omega^2+\textbf{k}^2+m_{\eta}^2
		-\frac{\Sigma_{\eta\mathrm{N}}}{2f_\pi^2}\rho^s_{N}
		+\frac{\kappa}{2f_\pi^2}\rho^s_{N}
		\left(-\omega^2+\textbf{k}^2\right)=0,
	\end{eqnarray}

	From the above equation, the  effective mass $m_{\eta}^*=\omega(| \textbf{k}|$=0) of $\eta$ meson can be written as
	\begin{eqnarray}
		m_{\eta}^*
		=\sqrt{
			\left(m_{\eta}^2-\frac{\Sigma_{\eta\mathrm{N}}}{2f_\pi^2}\rho^s_{N}\right)
			\Big/ \left(1+\frac{\kappa}{2f_\pi^2}\rho^s_{N}\right)
		}.
		\label{mseta}
	\end{eqnarray}
	
	Further, the $\eta$-meson self-energy  derived from Eq.(\ref{decom}) is given by
	\begin{eqnarray}
		\Pi^*(\omega,\textbf{k})
		=\Big (-\frac{\Sigma_{\eta\mathrm{N}}}{2f_\pi^2}
		+\frac{\kappa}{2f_\pi^2}
		(-\omega^2+\textbf{k}^2)\Big ) \rho^s_{N}.
	\end{eqnarray}

	
	\section{Results and Discussions}
	\label{sec:3}

	In this section,  at first we discuss the  behavior of in-medium nucleon scalar densities   in the hot asymmetric nuclear matter. Further, we discuss  the effective mass of $\eta$ meson which is  derived using the chiral SU(3) model alone in section  \ref{subsec:3.1}  and with the unified approach of ChPT and chiral SU(3) model  in section  \ref{subsec:3.2} . In  both approaches, we show the results for range of scattering length,  $a^{\eta_N}$=0.91-1.14 fm. Various parameters used in the present investigation are mentioned in \cref{ccc}.

\begin{figure}[h ]
		\includegraphics[width=16cm,height=16cm]{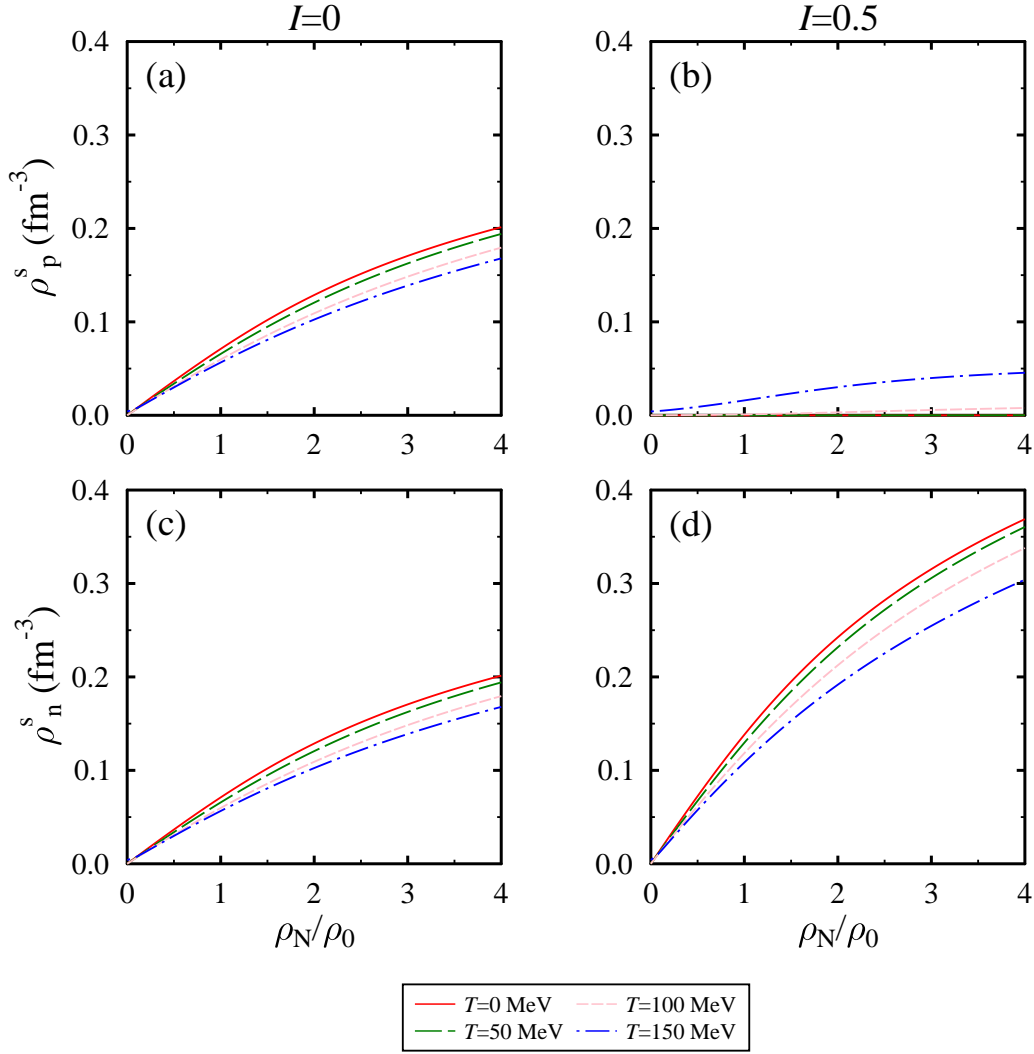}
		\caption{(Color online) The in-medium scalar density of nucleons. }
		\label{rs}
	\end{figure}

	In  chiral model, the scalar densities of nucleons have been calculated through Eq. (\ref{rhos0}). This equation contains the effect of medium modified scalar and vector fields \cite{Papazoglou1999}. The in-medium behavior of these fields is obtained by solving the coupled equations of motion (Eqs. (\ref{sigma}) to (\ref{chi})) \cite{Kumar2020b}.
	In \cref{rs}, we plot the scalar density of proton and neutron as a function of number density for finite values of temperature, $T$ and isospin asymmetry parameter. In  symmetric nuclear matter as  the contribution of $\delta$ and $\rho$ field is zero \cite{Kumar2019}, we get the same behavior of  neutron and proton scalar densities. The $\delta$ and $\rho$ field changes the in-medium value of baryon mass $m^*_i$ and effective chemical potential $\mu^*_i$, respectively which further modifies the nucleon scalar density (see Eq.(\ref{rhos0})) \cite{Papazoglou1999}.
	In figure, at $T=0$ the scalar density increases linearly in the low density regime  and  becomes non-linear in the high density regime. When we move from $I$=0  to \textit{I}$\neq$0 region, we observe a gradual increase in the neutron scalar density whereas the proton scalar density decreases. This is  because of  the non-zero contribution of $\delta$ and $\rho$ field in the isospin asymmetric nuclear matter which changes the effective mass as well as chemical potential and  therefore scalar density \cite{Papazoglou1999}.

	Another thermodynamic quantity $i.e.$  temperature is also a main property of the nuclear medium and in  \cref{rs} we have shown how the in-medium dynamics changes under non-zero temperature. The effect of temperature is observed more in the high density regime as compared to the low density regime. For symmetric matter in sub-plots (a) and (c), we anticipate  appreciable effect of  temperature. Here, for a particular value of nuclear density the value of scalar densities decrease as a function of  temperature. This is because of the Fermi distribution integral, due to the coupled nature of the equations (Eqs. (\ref{sigma}) to (\ref{chi})) the value of scalar density in Eq.(\ref{rhos0}) decreases when we increase the temperature in  integral. On the other hand,  in the highly asymmetric matter $i.e.$ $I$=0.5, for the neutron scalar density the temperature effects become more appreciable. In addition, we observe a minor  contribution to the proton scalar density for higher temperature values. This is because at finite temperature the proton condensate ($\bar p p$) $i.e.$ proton scalar density still populates in the medium despite the zero value of proton number density $\rho_p$. 
	The observed behavior of scalar densities in the symmetric nuclear matter is in agreement with the calculations of the Relativistic mean-field model \cite{Zhong2006,Song2008}. 
	

	\subsection{Optical Potential and Mass of $\eta$ Meson in Chiral Model}
	\label{subsec:3.1}
	

	 \begin{table}
\begin{tabular}{|c|c|c|c|c|c|c|c|c|c|}
\hline
& & \multicolumn{4}{c|}{$\eta$=0}    & \multicolumn{4}{c|}{$\eta$=0.5}   \\
\cline{3-10}
&$a^{\eta N} (\text{fm})$ & \multicolumn{2}{c|}{T=0} & \multicolumn{2}{c|}{T=100 }& \multicolumn{2}{c|}{T=0}& \multicolumn{2}{c|}{T=100 }\\
\cline{3-10}
&  &$\rho_0$&$4\rho_0$ &$\rho_0$  &$4\rho_0$ & $\rho_0$ &$4\rho_0$&$\rho_0$&$4\rho_0$ \\ \hline 
& 0.91&-46.18& -132.88& -37.78 &-120.31&-44.74&-123.35&-37.74&-116.46 \\ \cline{2-10}
$ \Delta m^*_\eta$&1.02&-54.61&-146.77  &-45.22 & -133.79 & -52.99 &-136.93&-45.16& -129.78 \\ \cline{2-10}
&1.14&-63.37&-160.51 &-52.98 & -147.21& -61.58 &-150.42 &-52.92& -143.07\\ \cline{1-10}

\end{tabular}
\caption{Values of in-medium mass-shift  of $\eta$-meson for different medium attributes calculated in chiral model are tabulated (in units of MeV).}
\label{tablems0}
\end{table}
	
\begin{figure}[h]
		\includegraphics[width=16cm,height=21cm]{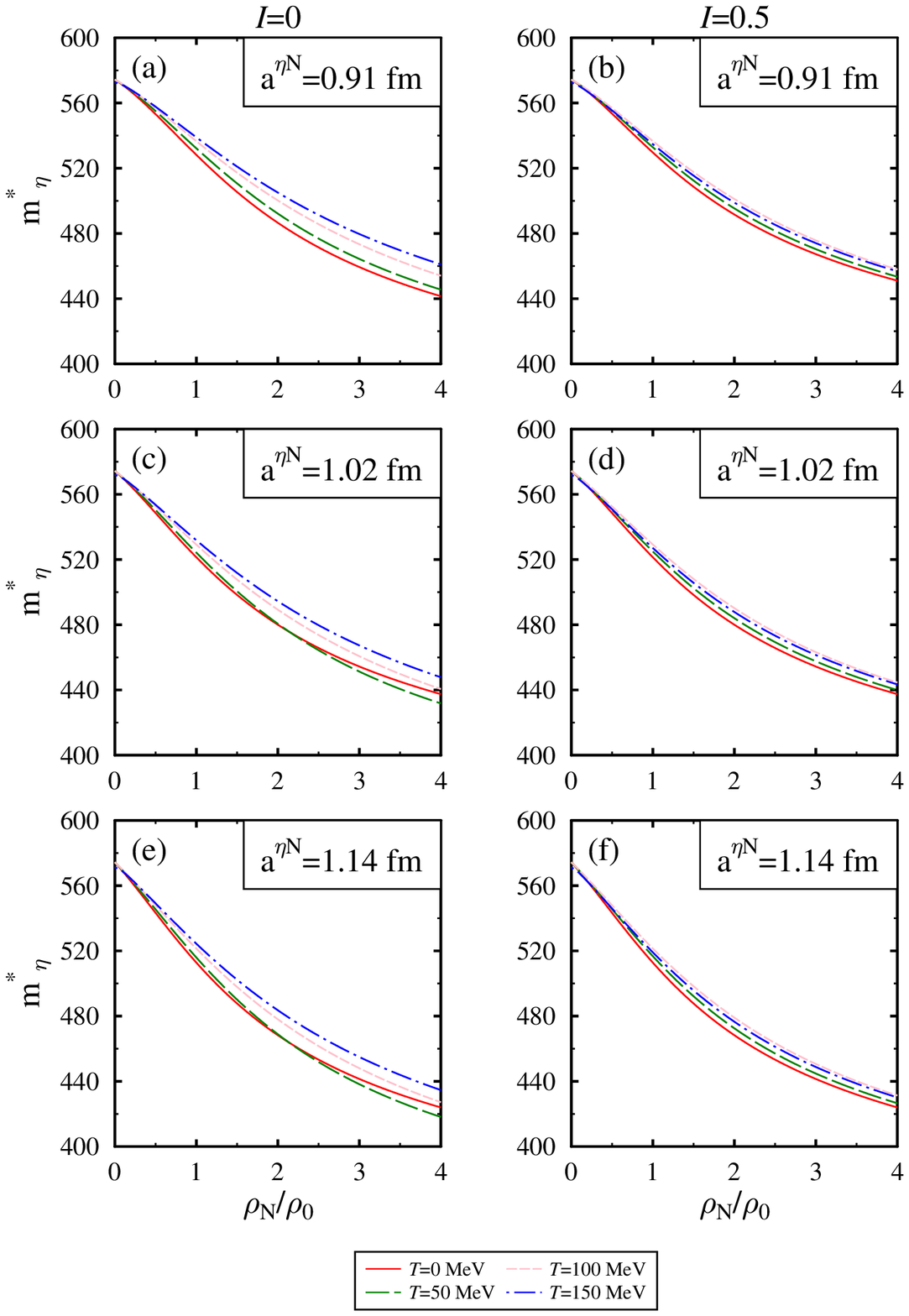}
		\caption{(Color online) In-medium $\eta$ meson mass in the chiral model. }
		\label{masseta}
	\end{figure}

	In \cref{masseta}, we have illustrated the medium modified mass of $\eta$-meson as a function of nuclear density for different values of scattering length. In the same figure, we also show the impact of isospin asymmetry and temperature. For given value of asymmetry, temperature,  scattering length,  the in-medium mass of $\eta$-meson is observed to decrease as a function of nuclear density. The rate of decrease is linear in the low density regime  whereas in the high density regime it becomes non-linear. This behavior  reflects the  opposite variation of nucleon scalar density plotted in \cref{rs}. This is because the  self-energy of $\eta$-meson  (see Eq.(\ref{sen})) has a direct dependence on the sum of scalar densities of nucleons.
	When we change the value of $a^{\eta N}$ from 0.91 to 1.14 fm, we observe a decrement in the effective mass. For example, at $\rho_N$=$\rho_0 (4\rho_0)$, $I$=$T$=0, the effective mass of $\eta$-meson changes from 528 (441) to 512 (423) MeV when we change    $a^{\eta N}$ value from 0.91 to 1.14 fm, respectively.  This is due to the $d^\prime$ term in  Eq.(\ref{sen}). The $d^\prime$ term has direct dependence on  $a^{\eta N}$ as shown in Eq.(\ref{dp}) and therefore  increasing the value of scattering length cause an increase in the value of $d^\prime$. Due to the attractive contribution of the self-energy  part corresponding to $d^\prime$ term the value of effective mass decreases. We also observed the substantial impact of the temperature on the in-medium mass in the symmetric nuclear matter which reflects the in-medium behavior of scalar densities. However, in the asymmetric nuclear matter, we observe the temperature effects on the mass to be less appreciable which reflects the less  contribution of the net scalar density ($\rho^s_p+\rho^s_n$).

\begin{figure}[h]
		\includegraphics[width=16cm,height=16cm]{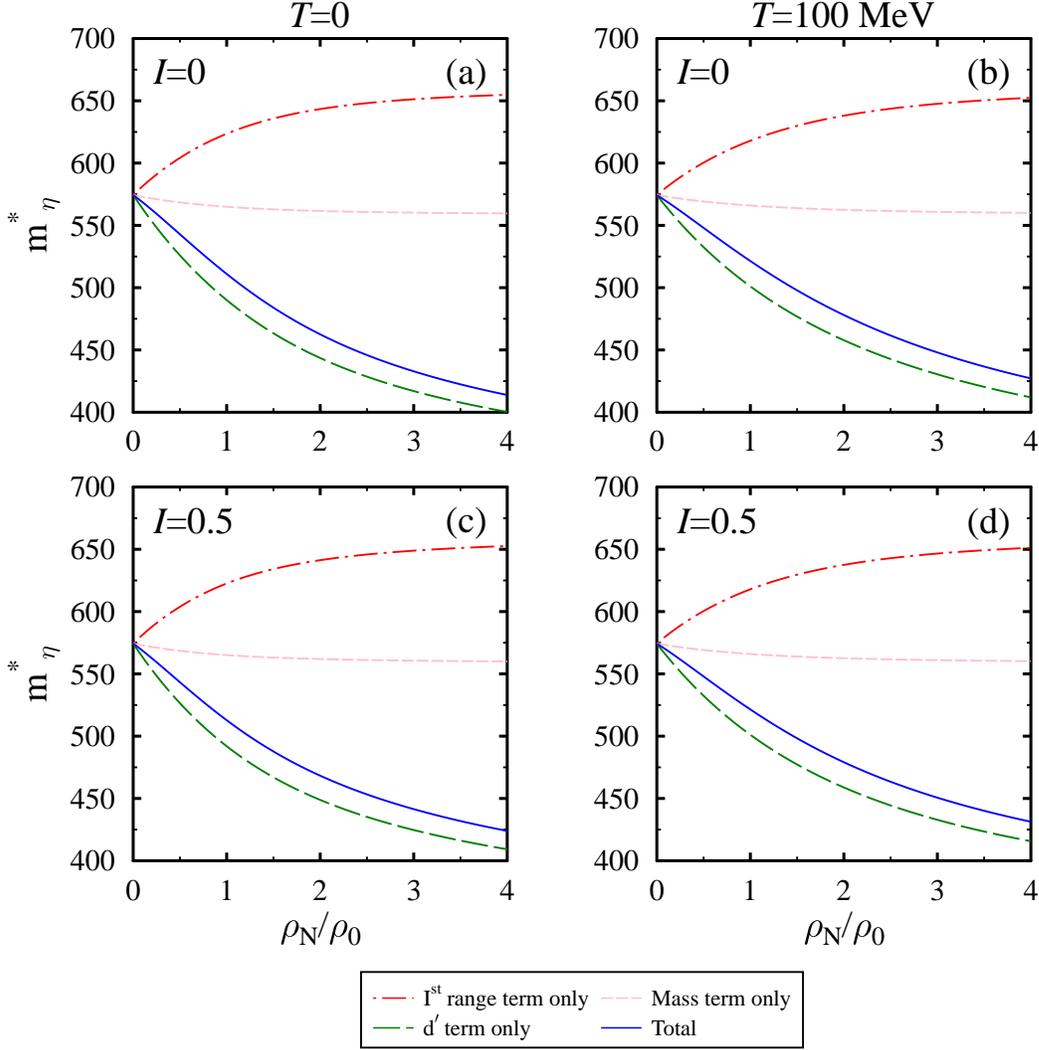}
		\caption{(Color online) Comparison of the different terms of  $\eta$-meson effective mass in chiral model at $a^{\eta N}=1.14$ fm. }
		\label{chiral_terms}
	\end{figure}

The self-energy  expression given by Eq.(\ref{sen}) contain three terms (i) first range term (ii) mass term and (iii) $d^\prime$ term. To understand the contribution of these individual terms, we illustrated   the in-medium mass of $\eta$-meson at zero and non-zero value of asymmetry and temperature in \cref{chiral_terms} for these different terms. At zero temperature and asymmetry,  one can see that the first range term gives an appreciable repulsive contribution to the effective mass whereas the mass and $d^\prime$ terms give attractive contributions.  We observe the dominant contribution of $d^\prime$ term which in turn gets reflected in the net effective mass. For non-zero temperature and asymmetry, the variation in the $d^\prime$ term becomes less hence we get a lower value of effective mass. This is due to the effect of scalar density terms present in the $d^\prime$ term (Eq.(\ref{sen})). 	For further  understanding, 
in \cref{masseta} we  plot the $\eta$-meson effective mass as a function of scattering length $a^{\eta N}$ at $\rho_N$=$\rho_0,4\rho_0$. At nuclear saturation density, we observe a linear decrease of effective mass with the increase in scattering length. Furthermore, the effective mass decrease more rapidly in the high density regime.
The observed behaviour emphasizes the importance of scattering length in the  $\eta N$ interactions.

\begin{figure}[h]
		\includegraphics[width=16cm,height=16cm]{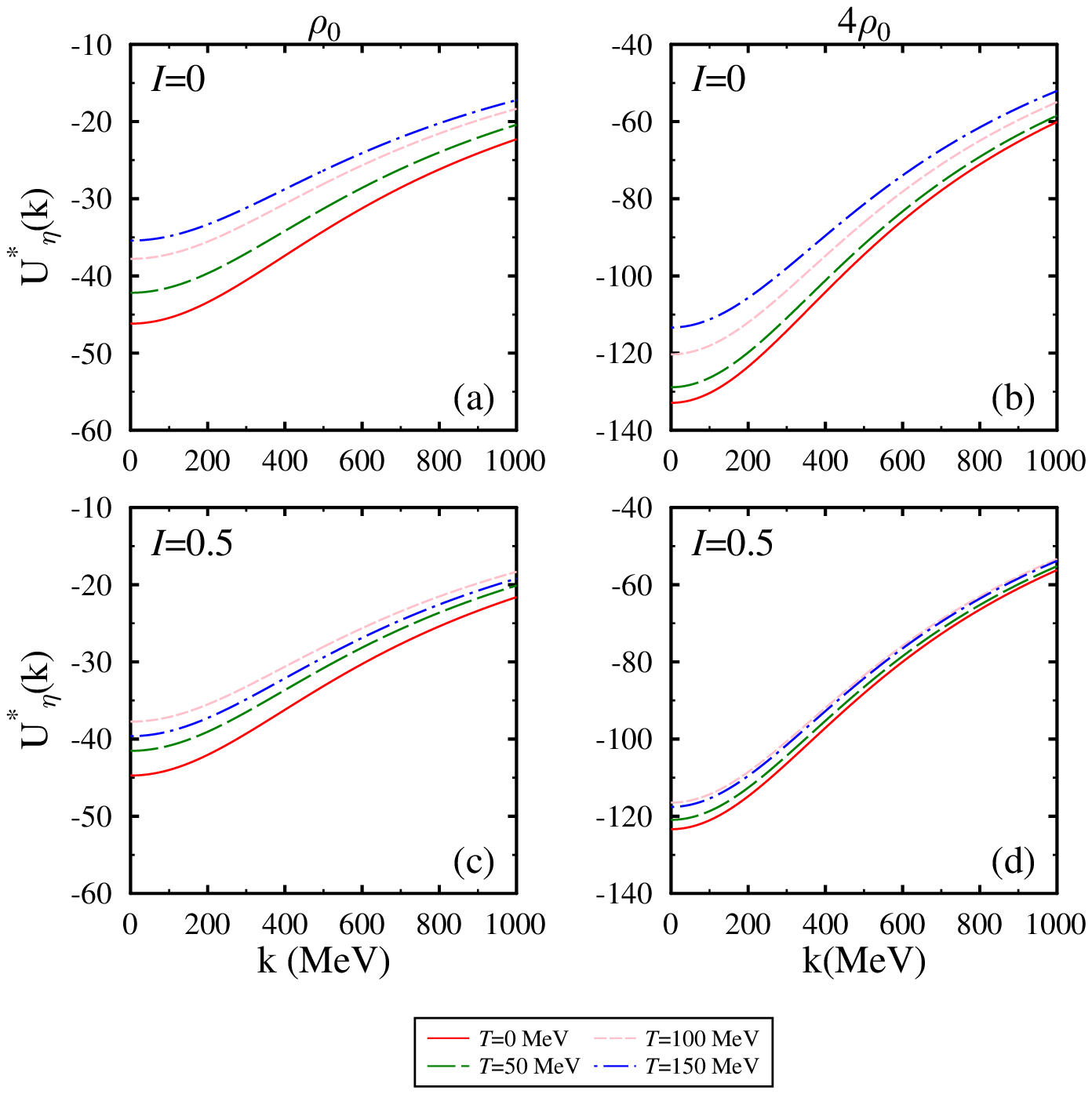}
		\caption{(Color online) The in-medium $\eta$ meson optical potential in chiral model  at $a^{\eta N}$=0.91 fm. }
		\label{Un0.91}
	\end{figure}

\begin{figure}[h]
		\includegraphics[width=16cm,height=16cm]{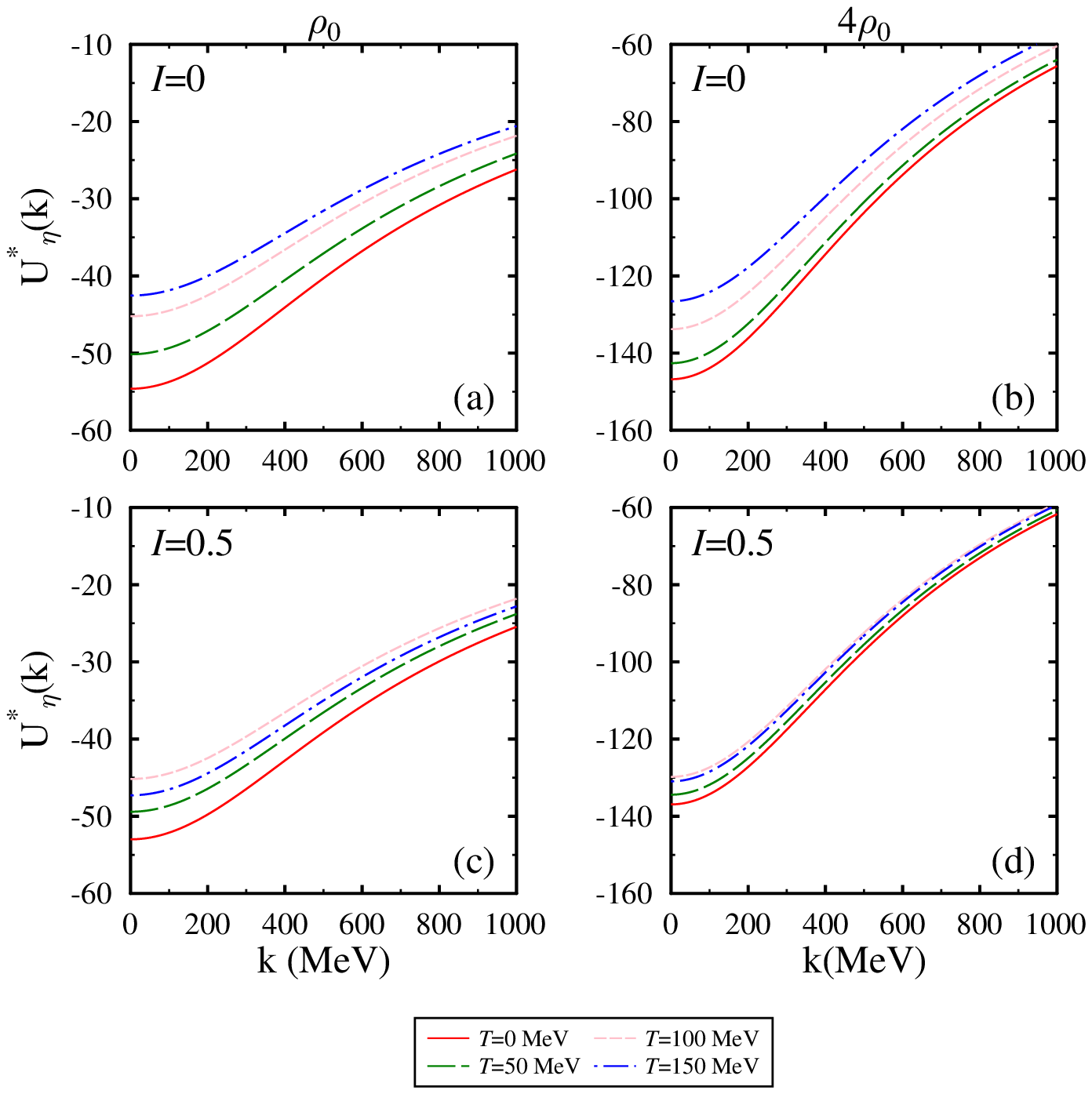}
		\caption{(Color online) The in-medium $\eta$ meson optical potential in chiral model  at $a^{\eta N}$=1.02 fm.}
		\label{Un1.02}
	\end{figure}

\begin{figure}[h]
		\includegraphics[width=16cm,height=16cm]{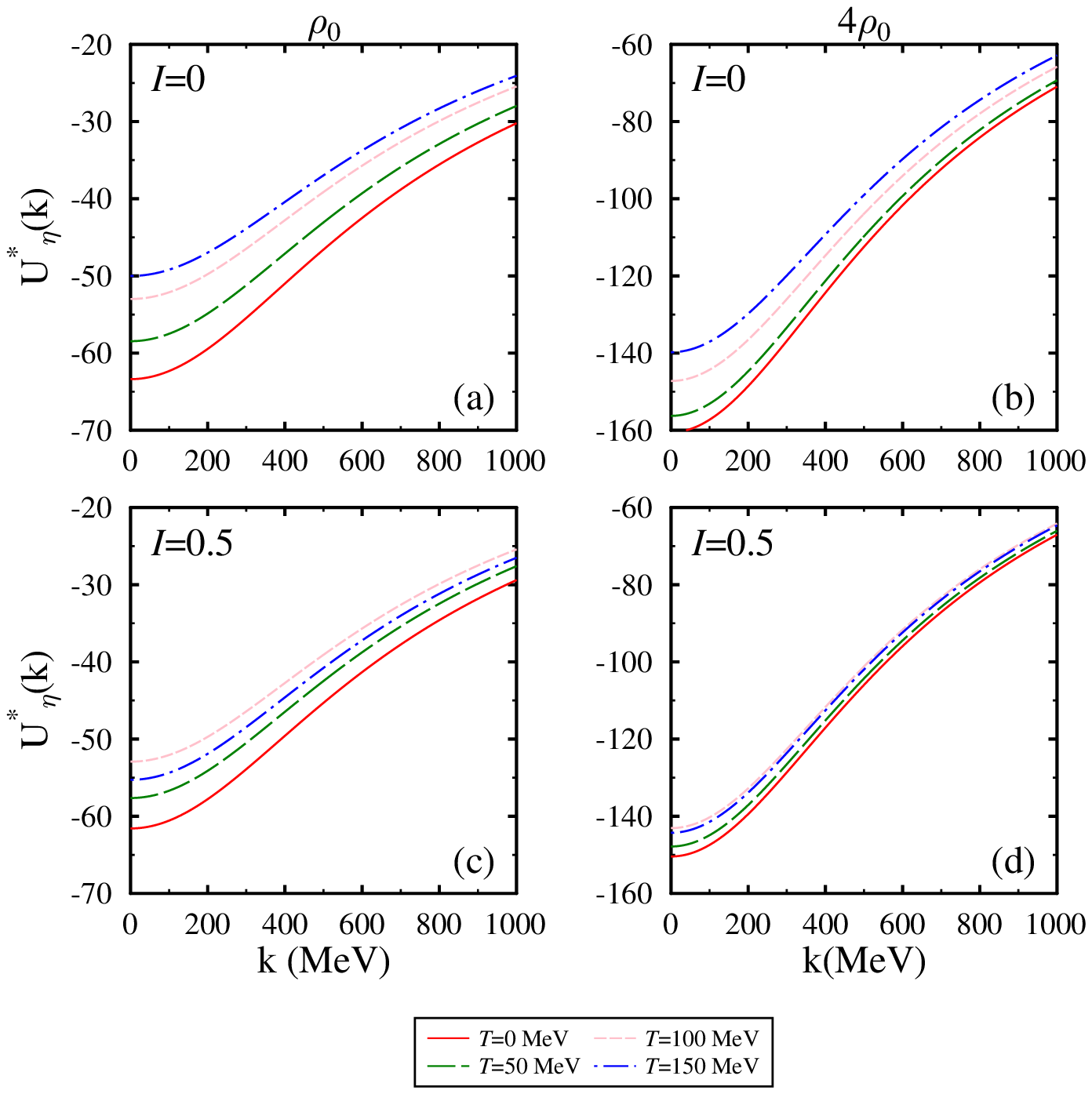}
		\caption{(Color online) The in-medium $\eta$ meson optical potential  in chiral model at $a^{\eta N}$=1.14 fm. }
		\label{Un1.14}
	\end{figure}

	The decrease in the  in-medium mass  leads to a negative  mass-shift which suggests the bound state formation of $\eta$-meson with a nucleus \cite{Jenkins1991,Zhong2006}. To understand the bound state phenomenon, the study of in-medium optical potential is very imperative. By using the effective mass in  Eq.(\ref{opc}), we plotted the optical potential of $\eta$-meson as a function of momentum $\lvert\textbf{k}\rvert$ for different values of $\eta N$ scattering length and other medium parameters  in \cref{Un0.91,Un1.02,Un1.14}. In \cref{Un0.91}, at $\rho_N$=$\rho_0$ we observe a negative value of  the optical potential. The value of optical potential becomes less negative as we increase the momentum of the $\eta$-meson.  The variation of optical potential reflects the interplay between the effective mass and momentum. At high values of the momentum,  Eq.(\ref{opc}) gets dominated by momentum and the contribution of effective mass becomes less.
	
	 A similar phenomenon happens in the high density regime. In this region, we observe  appreciable values of optical potential which become less as momentum  increases. Moreover, in the presence of a high density of neutron matter
	, we anticipate less effect of temperature which reflects the in-medium behavior of $\eta$-meson
	mass.  In \cref{Un1.02,Un1.14} we observe a similar trend of optical potential with $\eta$ momentum. In these figures, we observe a more negative value of optical potential as we increase the scattering length.  As discussed earlier, the optical potential is directly related to in-medium mass,  here it is illustrated to get a clear idea of negative potential. In the cold symmetric nuclear matter, at  $\rho_N$=$\rho_0(4\rho_0)$ we observe optical potential to be -54.61 (-146.77) MeV for $a^{\eta N}$=1.02 fm, whereas for $I$=0.5 these values changes to -52.99 (-136.93) MeV.  For better understanding, we have tabulated the in-medium mass-shift of $\eta$-meson at zero momentum in \cref{tablems0}.

	\begin{figure}[h]
		\includegraphics[width=16cm,height=16cm]{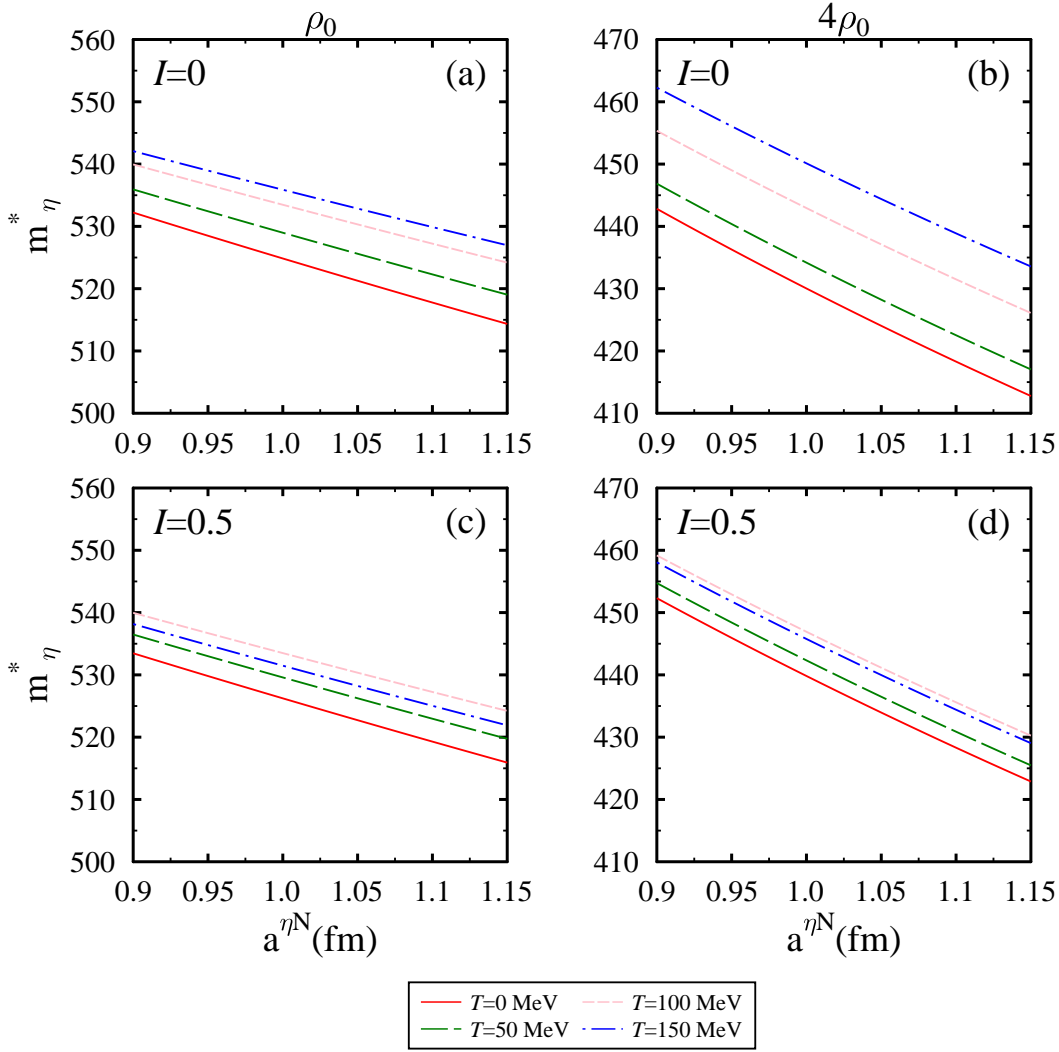}
		\caption{(Color online) The in-medium $\eta$-meson mass as a function of scattering length. }
		\label{massetavsa}
	\end{figure}

	%

	\subsection{In-medium Mass of $\eta$ Meson in Unified Approach of  ChPT and Chiral Model}
	\label{subsec:3.2}

	 \begin{table}
\begin{tabular}{|c|c|c|c|c|c|c|c|c|c|}
\hline
& & \multicolumn{4}{c|}{$\eta$=0}    & \multicolumn{4}{c|}{$\eta$=0.5}   \\
\cline{3-10}
&$a^{\eta N} (\text{fm})$ & \multicolumn{2}{c|}{T=0} & \multicolumn{2}{c|}{T=100 }& \multicolumn{2}{c|}{T=0}& \multicolumn{2}{c|}{T=100 }\\
\cline{3-10}
&  &$\rho_0$&$4\rho_0$ &$\rho_0$  &$4\rho_0$ & $\rho_0$ &$4\rho_0$&$\rho_0$&$4\rho_0$ \\ \hline 
& 0.91&-107.54& -219.71& -93.73 &-205.06&-105.19&-208.52&-93.64&-200.43 \\ \cline{2-10}
$ \Delta m^*_\eta$&1.02&-116.83&-232.28  &-102.21  & -217.49 & -114.35 &-220.99&-102.11 & -212.80 \\ \cline{2-10}
&1.14&-126.36&-244.56 &-110.96&-229.72 & -123.75 &-233.24& -110.86 &-225.00\\ \cline{1-10}

\end{tabular}
\caption{Values of in-medium mass-shift  of $\eta$-meson for different medium attributes calculated in the ChPT+chiral model are tabulated (in units of MeV).}
\label{tablems}
\end{table}

	In this section, we have used the unified approach of chiral SU(3) model and chiral perturbation theory to calculate the medium induced mass of $\eta$-meson. As discussed in the methodology section, the $\eta N$ equation of motion is obtained from the Lagrangian density of ChPT. Further, the scalar density of nucleons appearing in the ChPT equation of motion  is obtained from the chiral SU(3) model. In this calculation, we have  taken the value of parameter $\Sigma_{\eta N}$ to be 280 MeV. We have not considered the contribution of uncertainties in the  $\Sigma_{\eta N}$  parameter because of the less contribution of sigma term in  the  in-medium mass as compared to $\kappa$ term which we will see later. The value of in-medium $\eta$ mass-shift calculated using present unified approach are given in  \cref{tablems}.

\begin{figure}[h]
		\includegraphics[width=16cm,height=16cm]{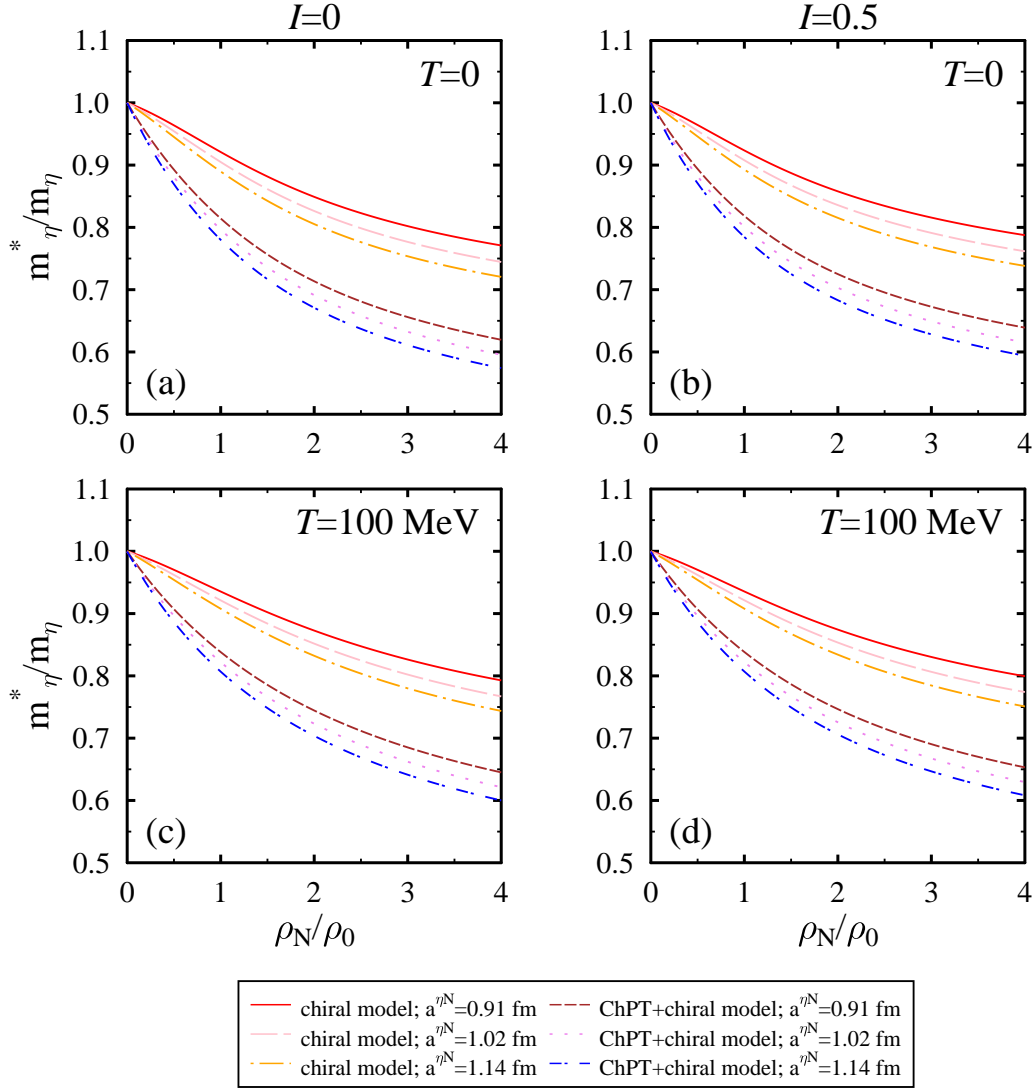}
		\caption{(Color online) Comparison of in-medium $\eta$ meson mass calculated from chiral model and ChPT. }
		\label{massra}
	\end{figure}

\begin{figure}[h]
		\includegraphics[width=16cm,height=16cm]{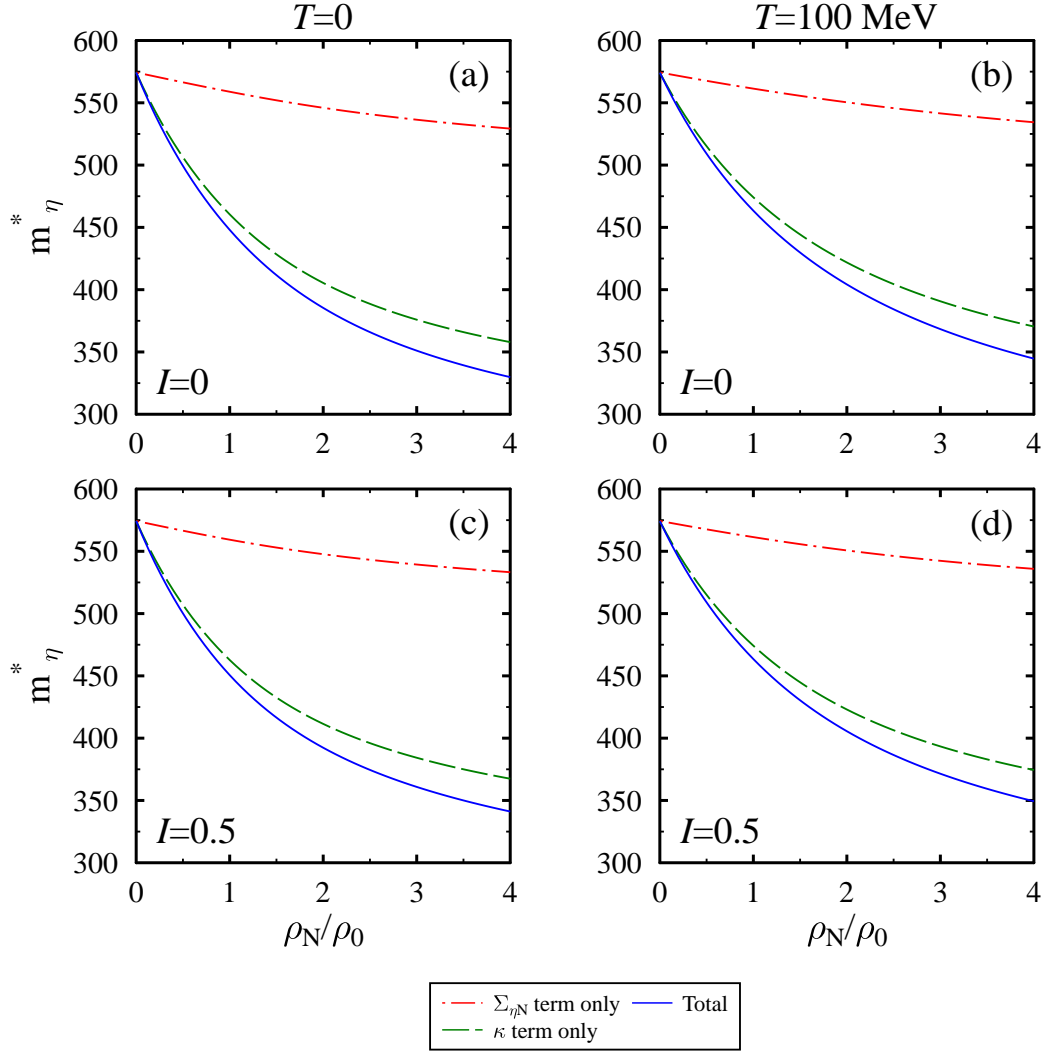}
		\caption{(Color online) Comparison of different terms of  the effective mass of $\eta$-meson calculated using unification of ChPT and chiral model at $a^{\eta N}=1.14$ fm. }
		\label{ChPT_terms}
	\end{figure}

	In \cref{massra}, we illustrated the ratio of  the in-medium  and vacuum mass of the $\eta$-meson as a function of nuclear density. In this figure, we have also included the effect of $\eta N$ scattering length, temperature, and medium isospin asymmetry. Moreover, we compared the results obtained from two different approaches $i.e.$ (i) chiral model alone (ii) ChPT and chiral model. Using the second approach,  we observed a substantial decrease in the mass of $\eta$-meson. We observed the same behavior of  the  in-medium mass with respect to temperature, asymmetry, and scattering length as was observed in the situation when the chiral model was used alone. The main difference is, in the ChPT the $\eta$
	meson gets a more net attractive contribution than the chiral model which is due to the absence of first range term  in ChPT Lagrangian. 	In \cref{ChPT_terms} we have  plotted the contributions of individual terms to the in-medium mass of $\eta$-meson calculated from the unified approach. The $\eta$-meson in-medium mass given by Eq.(\ref{mseta}) in the ChPT+chiral model  approach has two terms (i) $\Sigma_{\eta N}$ term and  (ii) $\kappa$ term. In this figure, we have shown the individual contribution of these  terms with increasing nuclear density and observed a non-appreciable contribution with $\Sigma_{\eta N}$ but appreciable with $\kappa$ term. This is because in the  $\eta$-meson in-medium mass expression (see Eq.(\ref{mseta})), the denominator  has a  positive contribution of  the scalar densities and the increase in scalar density with number density increases the denominator hence the value of effective mass becomes more negative. Clearly, there is no first range  term with the positive contribution as was observed in the previous chiral model calculations, and  therefore in the present case we get substantial attractive mass-shift.  
	
	The present observations can be compared with the $\eta$-meson effective mass calculated in the unified approach of ChPT and relativistic mean-field model  of Ref. \cite{Zhong2006}. In this article, authors also considered the effect of scattering length and at nuclear saturation density and $a^{\eta N}$=1.02 fm they anticipated the effective mass to be 0.84 $m_{\eta}$ whereas we observed it to  be 0.79 $m_{\eta}$.  At  nuclear saturation density, the effective mass equal to 0.95 $m_{\eta}$ was obtained within the coupled channel approach with scattering
	length $a^{\eta N}$ $\sim$ 0.25 fm \cite{Waas1997}.  In this non-diagonal coupled channel approach, there are only leading order contributions and hence, only a small decrement in the in-medium mass is observed. Also, in the QMC model at $\rho_0$ the in-medium mass of $\eta$-meson having value  0.88 $m_{\eta}$ was observed \cite{Tsushima1998}.  The obtained values are comparable with the calculations of the ChPT+chiral model for $a^{\eta N}$ $\sim$ 0.50 fm. 
	
	 In the cold symmetric nuclear matter, at  $\rho_N$=$\rho_0 (4\rho_0)$ and  $\lvert \textbf{k} \rvert$=0, we observe optical potential to be -116.83 (-232.28) MeV for $a^{\eta N}$=1.02 fm and in the cold isospin asymmetric nuclear matter the values modifies to -114.35 (-220.99) MeV. Using the ChPT+chiral model approach, we observed a even deeper optical potential than evaluated  in the  relativistic mean-field model+ChPT approach of Ref. \cite{Zhong2006}. This is due to the difference in the in-medium scalar densities obtained in two models. In our approach, we have taken the effect of scalar and vector fields under the influence of isospin asymmetry, and finite temperature whereas in the relativistic model approach only cold symmetric medium was considered. The  $\eta$ optical potential was also observed in the different theoretical observations \cite{Waas1997,Tsushima1998,Cieply2014,Inoue2002,Zhong2006}.   $U_{\eta}$=-34 MeV was observed by studying the $\eta N$ interactions near the threshold using free space chirally inspired coupled channel approach by considering the contributions of $N^*$(1535) baryon resonance \cite{Cieply2014}.  
	Besides, the optical potential $U_{\eta}$=-54, -60, and -83 MeV was observed in the  chiral unitary approach \cite{Inoue2002},  QMC model \cite{Tsushima1998}, and the ChPT \cite{Zhong2006}, respectively.
	

	\section{SUMMARY}
	\label{sec:4}

	We investigated the in-medium mass of $\eta$-meson in the asymmetric nucear matter at finite temperature. Under these medium conditions, we studied the behavior of the $\eta$-meson  using two different methodologies. In the first methodology, using the chiral model alone we calculated the medium modified mass and optical potential of $\eta$-meson by considering the $\eta N$ interactions up to second order in the Lagrangian and observe a   decrease in the effective mass of $\eta$-meson as a function of density. We find the in-medium effects to be more appreciable in the high density regime. In the second, we used the unified approach of chiral perturbation theory (ChPT) and chiral SU(3) model to study the in-medium attributes of $\eta$-meson. In this approach, we took the next-to-leading order contributions. We incorporated the medium effects from the chiral SU(3) model through scalar density which is plugged in the $\eta N$ equation of motion, which is calculated from the effective $\eta N$ Lagrangian of ChPT. Using this methodology, we find a substantial decrease in the mass of $\eta$-meson as a function of nuclear density.  The temperature and asymmetry effects are also studied and found to be slight repulsive in nature. Also, in the both approaches the mass-shift is observed to increase with an increase in the value of scattering length. The decrement on the $\eta$-meson mass leads to a negative mass-shift/optical potential which further suggests the possibility of $\eta N$ bound states.  The optical potential calculated in the present work will be used in future  to calculate the spectroscopic state of the $\eta$-mesic nuclei \cite{Zhong2006}.  Also, the momentum dependent optical potential can be used to study the $\eta$-meson production rate \cite{Peng1987,Martinez1999,Agakishiev2013} and  its momentum dependence in the asymmetric nuclear medium \cite{David2018,Chen2017,Berg1994}. 
	
	%
	%
	%
	%
	%
	%
	%
	%
	%
	%
	%
	%
	%

	\section*{Acknowledgment}

	One of the authors, (R.K)  sincerely acknowledge the support towards this work from Ministry of Science and Human Resources Development (MHRD), Government of India via Institute fellowship under National Institute of Technology Jalandhar.

	\appendix
	
	\section{EXPLICIT REPRESENTATION OF DIFFERENT  MATRICES}
	\label{appendix}
	Here, we give the matrix representation of meson, baryon and mass matrices which are used in present calculations \cite{Papazoglou1999}.
	\begin{itemize}
		\item{The Scalar Meson Matrix, $X$: }
		
		\be
		\label{smat}
		X=\frac{1}{\sqrt{2}}\sigma^a \lambda_a=
		\left( \begin{array}{ccc}
			(\delta  +\sigma)/\sqrt{2} & \delta^+ & \kappa^+\\   
			\delta^- & (-\delta+\sigma)/\sqrt{2} & \kappa^0 \\
			\kappa^- & \ovl{\kappa^0}& \zeta 
		\end{array} \right).
		\ee
		
		\item{The Pseudoscalar Meson Matrix, $P$:}
		
		\be
		P=\frac{1}{\sqrt{2}}\pi_a \lambda^a
		=\left (\begin {array}{ccc} 
		\frac{1}{\sqrt{2}}\left ( \pi^0+{\frac {\eta}{\sqrt {1+2\,{w}^{2}}}}\right )&\pi^{+}
		&2\,{\frac {K^{+}}{w+1}}\\
		\noalign{\medskip}\pi^{-}&\frac{1}{\sqrt{2}}\left 
		(- \pi^0+
		{\frac {\eta}{\sqrt {1+2\,{w}^{2}}}}\right )&2\,{\frac { K^0
			}{w+1}}\\\noalign{\medskip}2\,{\frac {K^-}{w+1}}&2\,{\frac { 
				\ovl{K}^0}{w+1}}&-{\frac {\eta\,\sqrt {2}}{\sqrt {1+2\,{w}^{2}}}}
		\end {array}\right ),
		\label{psmat}
		\ee
		
		where $w=\sqrt{2}\zeta_0/\sigma_0$.
		
		\item{The $A_p$ Matrix:}
		
		\begin{equation}
			\label{apmat}
			A_p=\frac{1}{\sqrt{2}}
			\left( \begin{array}{ccc}
				m_{\pi}^2 f_{\pi}& 0& 0\\   
				0 & m_\pi^2 f_\pi& 0\\
				0 & 0& 2 m_K^2 f_K
				-m_{\pi}^2 f_\pi
			\end{array} \right).
		\end{equation}
		
		\item{The Baryon Matrix, $B$: }
		\be
		\label{bmat}
		B=\frac{1}{\sqrt{2}}b^a \lambda_a=
		\left( \begin{array}{ccc}
			\frac{\Sigma^0}{\sqrt{2}} +\frac{\Lambda^0}{\sqrt{6}}& \Sigma^+ & p\\   
			\Sigma^- & -\frac{\Sigma^0}{\sqrt{2}} +\frac{\Lambda^0}{\sqrt{6}} & n \\
			\Xi^- & \Xi^0& -2 \frac{\Lambda^0}{\sqrt{6}}
		\end{array} \right).
		\ee

	\end{itemize}

\end{document}